\documentclass[12pt]{article}
\usepackage{cite}

\newcommand{\hoch}[1]{$\, ^{#1}$}
\newcommand{\fft}[2]{{\frac{#1}{#2}}}
\newcommand{\ft}[2]{{\textstyle\frac{#1}{#2}}}

\makeatletter
\@addtoreset{equation}{section}
\makeatother

\begin{document}

\begin{titlepage}

\begin{flushright}
CAMS/03-04\\
MCTP-03-18\\
hep-th/0304253
\end{flushright}

\vspace{15pt}

\begin{center}

{\LARGE Consistent reductions of IIB$^*$/M$^*$ theory and\\[8pt]
de~Sitter supergravity}

\vspace{15pt}

{James T.~Liu\hoch{*}, W.~A.~Sabra\hoch{\dagger} and W.~Y.~Wen\hoch{*,}%
\footnote{Email addresses: jimliu@umich.edu, ws00@aub.edu.lb,
wenw@umich.edu}}

\vspace{7pt}
\hoch{*}{Michigan Center for Theoretical Physics\\
Randall Laboratory, Department of Physics, University of Michigan\\
Ann Arbor, MI 48109--1120, USA}

\vspace{7pt}
\hoch{\dagger}{Center for Advanced Mathematical Sciences (CAMS) and\\
Physics Department, American University of Beirut, Lebanon}

\end{center}  

\begin{abstract}
We construct consistent non-linear Kaluza Klein reduction ansatze for
a subset of fields arising from the reduction of IIB$^*$ and M$^*$ theory
on dS$_5\times H^5$ and dS$_4\times$AdS$_7$, respectively.  These reductions
yield four and five-dimensional de Sitter supergravities, albeit with wrong
sign kinetic terms.  We also demonstrate that the ansatze may be used to
lift multi-centered de Sitter black hole solutions to ten and eleven
dimensions.  The lifted dS$_5$ black holes correspond to rotating E4-branes
of IIB$^*$ theory.
\end{abstract}

\end{titlepage}

\section{Introduction}

The recent observations on the acceleration of the universe has led to
renewed interest in de Sitter backgrounds in cosmology.  At the same
time, this has led to much debate on whether de Sitter space itself is
in fact compatible with M-theory.  Although several arguments may be
made against this possibility, it should be noted that de Sitter spaces
naturally arise as backgrounds for the * theories of
\cite{Hull:1998vg,Hull2,Hull3,Hull4,Hull}.  These * theories are obtained
through a timelike T-duality of the ordinary string theories, and admit an
unconventional effective field theory description involving wrong sign
kinetic energy terms for the RR fields.  Because of this and other
problems, the field theory limits of the * theories appear ill-defined.
Nevertheless, so long as one allows T-duality along a timelike circle,
one must allow for the existence of such * theories as a component of the
full M-theory.  It is in this spirit that we choose to investigate the
de Sitter supergravities which arise as consistent dimensional
reductions of IIB$^*$ and M$^*$ theory.

Conventional wisdom indicates that de Sitter space is
incompatible with supersymmetry.  This may be seen, for example, in
the standard classification of possible supersymmetry algebras, which allows
for both Poincar\'e and anti-de Sitter superalgebras, but not for the
possibility of a de Sitter superalgebra.  Thus in order to obtain de
Sitter supersymmetry, one must relax one or more of the usual assumptions.
Such possibilities were first generally studied in \cite{Lukierski,Pilch}.
In the present case, the de Sitter supergravities which we investigate
are unconventional in that they have wrong sign kinetic energy terms
as well as non-compact gaugings.  Of course, unitary theories do exist
with non-compact gaugings.  However, as will be explained below, the
non-compact symmetries which we consider here are of a different
nature, and yield mixed sign kinetic terms for the gauge fields.
While these properties of the de Sitter supergravities are clearly
undesirable, they follow as a natural consequence of the underlying *
theory.  As a result, we leave open the possibility that such de Sitter
supergravities would be of relevance for the * theory beyond the
field theory limit.  It is, however, possible that the effective field
theories which we explore, being unstable, do not provide an adequate
description of the full * theory.  Nevertheless, one may hope to gain
additional insight into de Sitter space and supersymmetry regardless of
the ultimate fate of the * theories.

For the case of anti-de Sitter supersymmetry, it has long been known
that a generalized Freund-Rubin compactification yields backgrounds of
the form AdS$\times$Sphere.  Furthermore, a linearized Kaluza-Klein
analysis indicates that the zero mode fluctuations about the
AdS$\times$Sphere background gives rise to a maximal gauged supergravity
in the lower dimensional AdS space.  More recently, various full and
truncated non-linear Kaluza-Klein reductions have been constructed,
demonstrating the consistent embedding of the corresponding gauged
supergravities in the higher dimensional theory
\cite{deWit,Cvetic,Nastase:1999cb,Nastase,Lu:1999bc,Lu:1999bw,Cvetic:1999au,Cvetic:1999xx,Cvetic:2000eb,Cvetic2,Cvetic:2000dm}.
Based on the observation that such gauged supergravities yield a
negative cosmological constant of the form $\Lambda\sim-g^2$, one may at
least formally obtain a de Sitter supergravity through the analytic
continuation to imaginary coupling constant, namely $g\to i\hat g$.
While we follow this approach in spirit, it is important to note that
the de Sitter supergravities discussed here (being descended from *
theories) have only real bosonic fields and real gauge couplings
(although Majorana conditions on the fermions may have to be relaxed).

The non-linear Kaluza-Klein ansatze for the conventional sphere reductions
allow the embedding of various lower-dimensional solutions into the
underlying higher-dimensional theories.  For example, R-charged AdS black
holes may be lifted to ten and eleven dimensions, where they take on the
nature of rotating branes \cite{Cvetic}.  Similarly, the consistent
reductions constructed below allow us to lift dS black holes into the
original * theory.  In particular, multi-centered dS black holes have been
constructed in \cite{Kastor:1992nn,London}; these were furthermore
shown to satisfy an unconventional supersymmetry involving the imaginary
coupling constant mentioned above \cite{London,Liu}.  Here we demonstrate
that such dS black holes provide an interesting cosmological background for
* theory and furthermore investigate their lifting to ten or eleven dimensions.
Such multi-centered black holes have also been considered recently by
Behrndt and Cveti\v{c} as examples of time-dependent backgrounds in the
analytically continued de Sitter supergravity \cite{Behrndt}.

In the following section we describe the general procedure of obtaining
consistent reduction ansatze for * theories.  Then in sections 3 and 4
we turn to the specific cases of IIB$^*$ and M$^*$ reductions,
respectively.  The latter model is particularly interesting, as it
admits a dS$_4\times$AdS$_7$ background, whereupon either dS$_4$ or
AdS$_7$ may be viewed as the lower-dimensional spacetime.  Since both
cases arise from the same eleven-dimensional background, this hints at
some form of a dS$_4$/AdS$_7$ duality \cite{Batrachenko:2002pu}.  Finally,
we consider the
lifting of dS black holes in section 5, and conclude in section 6.

\section{Generalized sphere reductions of IIB$^*$ and M$^*$ theory}

It has been an important observation that non-dilatonic branes, in
particular the D3, M2 and M5 branes, serve as interpolating solutions
between asymptotic Minkowski and near horizon AdS$\times$Sphere
geometries (AdS$_5\times S^5$, AdS$_4\times S^7$ and AdS$_7\times S^4$,
respectively).  Alternatively, the AdS$\times$Sphere geometry directly
arises from a generalized Freund-Rubin compactification.  In several
cases, the complete non-linear Kaluza-Klein reduction corresponding to
such geometries is known.  However, it is generally more common that only
a truncated ansatz (often to a maximal Abelian subgroup of the full gauge
group, or with a subset of scalars and higher-rank potentials) has been
constructed.  Such truncated ansatze are often sufficient for the lifting
of solutions such as AdS black holes to the higher dimensional
theory.  

While it would be desirable to obtain a complete reduction, in this
paper we exclusively focus on the truncation to the sector arising from
the higher dimensional metric and $p$-form fields.  To motivate our
approach, consider the Kaluza-Klein sphere reduction ansatz to
AdS$_d\times S^n$.  In this case, the metric has the
general form \cite{Cvetic:1999xx,Cvetic:2000eb,Cvetic2}
\begin{equation}
ds_D^2=\Delta^{2/(d-1)}ds_d^2+g^{-2}\Delta^{-(d-3)/(d-1)}(T^{-1})^{ij}
D\mu^iD\mu^j.
\label{eq:metans}
\end{equation}
Here, $i$ and $j$ run from $1$ to $n+1$, and the $\mu^i$'s satisfy
the constraint $\sum_i\mu^{i\,2}=1$, corresponding to a parametrization
of $S^n$.  The $n$-sphere itself is given by the coset space
$SO(n+1)/SO(n)$, while $T_{ij}$ is a symmetric unimodular matrix
consisting of $\fft12n(n+3)$ scalar degrees of freedom parameterizing the
coset $SL(n+1,R)/SO(n+1)$.  The isometry of $S^n$ gives rise to a
$SO(n+1)$ gauge symmetry, with gauge potentials $A_{(1)}^{[ij]}$.  The
gauge covariant derivative is given by, {\it e.g.},
$D\mu^i=d\mu^i+gA_{(1)}^{ij}\mu^j$.  Associated with the metric ansatz
(\ref{eq:metans}) is a corresponding one on the $p$-form potential
($F_{4}$ for $D=11$, or $F_{5}$ for IIB).  This will be considered in
more detail below, when we specialize to the various cases at hand.

The above analysis of the near-horizon dynamics of non-dilatonic branes
may be generalized to encompass the * theories of \cite{Hull:1998vg,Hull2}.
For example, the IIA$^*$ and IIB$^*$ theories admit branes which are the
timelike T-duals of ordinary D branes, while M$^*$ theory admits
generalized M2 and M5 branes, all of which may have unusual signatures
on their world sheets \cite{Hull3,Hull4}.  In themselves, these branes
are all legitimate solutions of the * theories.  However, it has been
noted that they may be obtained from the ordinary brane solutions via
appropriate analytic continuations
\footnote{That this is the case simply follows from the fact that the
supergravity description of the * theories may be obtained by suitable
Wick rotations and analytic continuations of the usual supergravities.}.
This fact will be important to us below in constructing the generalized
sphere reductions.

Just as the ordinary non-dilatonic branes serve as interpolating
solutions between maximally symmetric spaces, the branes of * theory
serve a similar role.  However, in this case, the near horizon limits
are generalizations of the AdS$\times$Sphere geometries to different
signatures and different signs of the spacetime and internal space
curvatures.  In all such cases, the resulting geometries for either
spacetime or the internal manifold have the maximally symmetric coset
form $SO(s+1,t)/SO(s,t)$ or $SO(s,t+1)/SO(s,t)$ where $s$ and $t$ denote
space and time dimensions, respectively \cite{Hull4}.  It should be
noted that the internal spaces are often non-compact, and may include
time-like coordinates.  In particular, M$^*$ theory admits
an interesting dS$_4\times$AdS$_7$ vacuum, which may be obtained as the
near horizon of either a M2 $(3,0,-)$ or a M5 $(5,1,+)$ brane.  This has
led to speculation on a possible AdS/CFT duality between the worldvolume
theories of M2 and M5, with the roles of spacetime and internal space
interchanged \cite{Batrachenko:2002pu}.

In order to obtain the non-linear Kaluza-Klein ansatz for reduction on
either $SO(s+1,t)/SO(s,t)$ or $SO(s,t+1)/SO(s,t)$, we may analytically
continue away from the sphere, $S^n$, corresponding to $SO(n+1)/SO(n)$.
More specifically, starting with the homogeneous embedding of $S^n$ in
$R^{n+1}$, given by $(\mu^1)^2+(\mu^2)^2+\cdots+(\mu^{n+1})^2=1$, we
analytically continue an appropriate subset of $\mu^i$'s, namely%
\footnote{Throughout this paper, we use a caret to distinguish quantities
involved in the generalized reduction from those of the usual sphere
reduction.}
$\mu^i\to i\hat\mu^i$, while at the same time changing the signature of
the embedding space in the natural manner.  This generalization to a
non-compact internal space is conveniently encoded in terms of the
Lorentzian metric, $\eta_{ij}$, on the embedding space, with the
hyperbolic embedding specified by the constraint
$\eta_{ij}\hat\mu^i\hat\mu^j=-1$.  In this case, the generalization of
the metric reduction ansatz, (\ref{eq:metans}), takes the essentially
identical form
\begin{equation}
ds_D^2=\Delta^{2/(d-1)}ds_d^2+\hat g^{-2}\Delta^{-(d-3)/(d-1)}
\eta_{ij}(\hat T^{-1})^{jk}\eta_{kl} D\hat\mu^iD\hat\mu^l.
\label{eq:hatmetans}
\end{equation}
Note, however, that while the scalars are still represented by a
symmetric matrix $\hat T_{ij}$, they now parametrize the coset
$SL(s+1+t,R)/SO(s+1,t)$ or $SL(s+t+1,R)/SO(s,t+1)$ as appropriate to a
generalized signature internal space.  As a result, $\hat T_{ij}$ may
have negative eigenvalues, and $\hat T_{ij}=\eta_{ij}$ corresponds to
the vacuum with no scalar excitations.  

At this point, it is important to realize that our claim of negative
eigenvalues for $\hat T_{ij}$ yields an unconventional version of non-compact
gauging.  In an ordinary gauged supergravity theory, one may choose to
gauge a non-compact group, say, $SO(p,q)$.  Nevertheless, all fields
have conventional kinetic energies, and the theory remains unitary.  In
particular, the equivalent of $\hat T_{ij}$ for the conventional
supergravity never has negative eigenvalues, regardless of the compact
versus non-compact nature of the gauging.  Furthermore, when the gauging
is removed by taking $g\to0$, one recovers the standard ungauged
supergravity.  For a conventional non-compact gauging, the gauge group is
spontaneously broken to its maximal compact subgroup.  In this case, it
has been shown in \cite{HullWarner,Gibbons} that such theories arise from
dimensional-reductions on non-compact spaces, and that they may be
obtained via analytical continuation in the same spirit that we are
advocating here.  The difference with the present case, however, is that
here we choose to reduce the * theory on the maximally symmetric vacuum.
In particular, this means that, even in the ungauged limit, we retain
both compact and non-compact gauge fields (with both positive and negative
kinetic terms, related to the eigenvalues of $\eta_{ij}$).  In the
framework of \cite{HullWarner}, we would be expanding about an unstable
vacuum.  However we leave this issue as one that must be resolved by the
underlying * theory.

While (\ref{eq:hatmetans}) is the universal form of the metric ansatz
for a generalized internal space, we have at the moment left the nature
of the $p$-form field unspecified.  To proceed, we must specialize to
a particular theory.  This is what we carry out in the next section
(for IIB$^*$ theory) and the subsequent one (for M$^*$ theory).

\section{The $H^5$ reduction of IIB$^*$ supergravity}

The bosonic sector of IIB supergravity consists of the metric, dilaton
and 3-form field strength $H_{(3)}$ in the NSNS sector, as well as
$F_{(1)}$, $F_{(3)}$ and $F_{(5)}$ in the RR sector.  The complete
$S^5$ reduction of type IIB supergravity gives rise to five-dimensional
$SO(6)$ gauged $N=8$ supergravity.  However, by considering the
truncation of IIB supergravity to only the metric and self-dual 5-form,
one ends up in five dimensions with a truncation of $N=8$ supergravity
to a bosonic system with $SO(6)$ gauge fields and 20 scalars.  The
consistent $S^5$ reduction in this subsector (which is the one most
directly relevant to the D3-brane) was given in \cite{Cvetic2}, and has
the form
\begin{eqnarray}
ds_{10}^2&=&\Delta^{1/2}ds_5^2+g^{-2}\Delta^{-1/2}(T^{-1})^{ij}D\mu^iD\mu^j,
\nonumber\\
F_{(5)}&=&G_{(5)}+\ast G_{(5)},\nonumber\\
G_{(5)}&=&-gU\epsilon_{(5)}+g^{-1}((T^{-1})^{ij}\bar{\ast} DT_{jk})
\wedge(\mu^kD\mu^i)
-\frac{1}{2}g^{-2}(T^{-1})^{ik}(T^{-1})^{jl}\bar{\ast} F^{ij}_{(2)}
\wedge D\mu^k \wedge D\mu^l,\nonumber\\
\label{eq:ads5s5}
\end{eqnarray}
where
\begin{equation}
U=2T_{ij}T_{jk} \mu^i \mu^k -\Delta T_{ii},\qquad
\Delta=T_{ij}\mu^i \mu^j.
\end{equation}
Here, $\epsilon_{(5)}$ is the volume 5-form in the spacetime.
As usual, the field strength $F_{(2)}^{[ij]}$ and gauge covariant
derivatives are given in terms of the $SO(6)$ gauge fields $A_{(1)}^{[ij]}$
by
\begin{eqnarray}
F^{ij}_{(2)}&=&dA^{ij}_{(1)}+gA^{ik}_{(1)}\wedge A^{kj}_{(1)},\nonumber\\
DT_{ij}&=&dT_{ij}+gA^{ik}_{(1)}T_{kj}+gA^{jk}_{(1)}T_{ik},\nonumber\\
D\mu^i&=&d\mu^i + gA^{ij}_{(1)}\mu^j.
\label{eq:Ddef}
\end{eqnarray}
We mostly follow the notation of \cite{Cvetic2}, except that we
reserve the caret to denote quantities relevant to * theory.
The coordinates, $\mu^i$, are subject to the constraint
$\delta_{ij}\mu^i\mu^j=1$, and parametrize the internal $S^5$.

We now recall that IIB$^*$ supergravity with signature $(9,1)$ may be
obtained from IIB theory by changing the signs of the RR kinetic terms.
In particular, the wrong-sign kinetic term for the self-dual 5-form
gives rise to the ten-dimensional Einstein equation
\begin{equation}
R_{MN}=-\frac{1}{2\cdot2\cdot4!}\hat F_{MPQRS}\hat F_N{}^{PQRS},
\end{equation}
which in turn leads to a dS$_5\times H^5$ Freund-Rubin vacuum.  This simply
corresponds to an interchange of positive and negative curvatures
between spacetime and internal space (as is evident from the extra sign in
the above Einstein equation).  It should now be evident how the
appropriate IIB$^*$ reduction ansatz may be obtained by the analytic
continuation of (\ref{eq:ads5s5}).  To generate the wrong-sign kinetic
term, we take $F_{(5)}\to i\hat F_{(5)}$.  However, since
$G_{(5)}=-gU\epsilon_{(5)}+\cdots$, we avoid a resulting imaginary
$\hat G_{(5)}$ by simultaneously taking $g\to i\hat g$.  At this stage,
we find
\begin{eqnarray}
ds_{10}^2&=&\Delta^{1/2}ds_5^2-\hat g^{-2}\Delta^{-1/2}(T^{-1})^{ij}
D\mu^iD\mu^j,\nonumber\\
\hat G_{(5)}&=&-\hat gU\epsilon_{(5)}-\hat g^{-1}((T^{-1})^{ij}
\bar{\ast}DT_{jk}) \wedge(\mu^kD\mu^i)
-\frac{i}{2}\hat g^{-2}(T^{-1})^{ik}(T^{-1})^{jl}\bar{\ast} F^{ij}_{(2)}
\wedge D\mu^k \wedge D\mu^l.\nonumber\\
\end{eqnarray}
Here, we implicitly assume that both $U$ and $\Delta$ remain real
quantities throughout the analytic continuation.  This intermediate
result for the metric and 5-form is unsatisfactory in that the internal
five-dimensional space has the wrong signature and that $\hat G_{(5)}$ is
still complex because of the factor of $i$ in the last term.  Both
issues may be solved by the analytic continuation
\begin{eqnarray}
&&\mu^i\to i\hat \mu^i,\qquad\hphantom{A-A}\!\!\mu^6\to\hat\mu^6,\nonumber\\
&&A_{(1)}^{ij}\to -i\hat A_{(1)}^{ij},\qquad
A_{(1)}^{[i6]}\to -\hat A_{(1)}^{[i6]}\qquad (i,j=1,\dots,5).
\end{eqnarray}
It is this step that changes the $\mu^i$'s from parameterizing $S^5$ as
$SO(6)/SO(5)$ to parameterizing $H^5$ as $SO(5,1)/SO(5)$.  At the same
time, the analytic continuation of the gauge fields leads to the
non-compact gauge group $SO(5,1)$.

In order for $U$ and $\Delta$ to be real, we finally continue
appropriate entries in the $T_{ij}$ matrix, namely
\begin{equation}
T_{ij}\to \hat T_{ij},\qquad
T_{(i6)}\to i\hat T_{(i6)},\qquad
T_{66}\to-\hat T_{66} \qquad (i,j=1, \ldots,5),
\end{equation}
with corresponding inverse
\begin{eqnarray}
&&(T^{-1})^{ij}\to (\hat T^{-1})^{ij},\qquad
(T^{-1})^{(i6)}\to -i(\hat T^{-1})^{(i6)},\qquad
(T^{-1})^{66}\to-(\hat T^{-1})^{66}\nonumber\\
&&\kern26em (i,j=1, \ldots,5).
\end{eqnarray}
Note that $\hat T$ has a single negative eigenvalue, and $\det\hat T=-1$.
We reiterate that this negative eigenvalue results in both correct and
wrong-sign gauge kinetic energy terms showing up, depending on the
non-compact versus compact nature of the corresponding generator.
The resulting reduction ansatz for IIB$^*$ theory on dS$_5\times H^5$ is
most conveniently given in terms of the $SO(5,1)$ metric
$\eta_{ij}={\rm diag}(+,+,+,+,+,-)$.  We find
\begin{eqnarray}
ds^2_{10}&=&\Delta^{1/2}ds_5^2+\hat{g}^{-2}\Delta^{-1/2}
\eta_{ij}(\hat T^{-1})^{jk}\eta_{kl}D\hat\mu^{i}D\hat\mu^{l},\nonumber\\
\hat{F}_{(5)}&=&\hat{G}_{(5)}+\ast \hat{G}_{(5)},\nonumber\\
\hat{G}_{(5)}&=&-\hat{g}U\epsilon_{(5)}
+\hat{g}^{-1}(\eta_{ij}(\hat T^{-1})^{jk}\bar{\ast}D\hat T_{kl})\wedge
(\hat{\mu}^{l}D\hat\mu^{i})\nonumber\\
&&\qquad-\frac{1}{2}\hat{g}^{-2}\eta_{ij}(\hat T^{-1})^{jk}\eta_{kl}
\eta_{mn}(\hat T^{-1})^{np}\eta_{pq}\bar{\ast}
\hat{F}^{im}_{(2)} \wedge (D\hat\mu^{l} \wedge D\hat\mu^{q}),
\label{eq:b5redans}
\end{eqnarray}
where
\begin{equation}
U=-2\hat T_{ij}\eta^{jk}\hat T_{kl} \hat{\mu}^i \hat{\mu}^l
-\Delta \eta^{ij}\hat T_{ij},\qquad
\Delta=-\hat T_{ij} \hat{\mu}^i \hat{\mu}^j,\qquad
\eta_{ij}\hat{\mu}^i \hat{\mu}^j = -1,
\end{equation}
and the gauge covariant derivatives are given by
\begin{eqnarray}
\hat{F}^{ij}_{(2)}&=&d\hat{A}^{ij}_{(1)}
+\hat{g}\eta_{kl}\hat{A}^{ik}_{(1)}\wedge\hat{A}^{lj}_{(1)},\nonumber\\
D\hat T_{ij}&=&d\hat T_{ij}+\hat{g}\eta_{ik}\hat{A}^{kl}_{(1)}\hat T_{lj}
+\hat{g}\eta_{jk}\hat{A}^{kl}_{(1)}\hat T_{il},\nonumber\\
D\hat\mu^i&=&d\hat{\mu}^i + \hat{g}\eta_{jk}\hat{A}^{ij}_{(1)}\hat{\mu}^k.
\label{eq:nccovar}
\end{eqnarray}
Although the structure of the non-compact gauging is perhaps obvious in
this reduction, to avoid confusion we will always retain explicit
factors of the $SO(5,1)$ metric $\eta_{ij}$.

Since this reduction of IIB$^*$ supergravity on dS$_5\times H^5$ was
obtained by appropriate continuation of the ansatz given in
\cite{Cvetic2}, it follows that the resulting five-dimensional
Lagrangian may similarly be obtained through analytic continuation.  The
resulting Lagrangian has the form \cite{Cvetic2}
\begin{eqnarray}
{\cal L}_5&=&R\,\bar{\ast}1-\frac{1}{4}(\hat T^{-1})^{ij}\bar{\ast}D\hat
T_{jk}\wedge (\hat T^{-1})^{kl}D\hat T_{li}
+\frac{1}{4}\eta_{ij}(\hat T^{-1})^{jk}\eta_{kl}
\eta_{mn}(\hat T^{-1})^{np}\eta_{pq}
\bar{\ast}\hat F^{im}_{(2)}\wedge\hat F^{lq}_{(2)}\nonumber\\
&&\qquad-V\,\bar{\ast}1
-\frac{1}{48}\epsilon_{i_1i_2i_3i_4i_5i_6}\Bigl(\hat F^{i_1i_2}_{(2)}
\hat F^{i_3i_4}_{(2)}\hat A^{i_5i_6}_{(1)}-\hat g\hat F^{i_1i_2}_{(2)}
\hat A^{i_3i_4}_{(1)}\hat A^{i_5j}_{(1)}\eta_{jk}\hat A^{ki_6}_{(1)} 
\nonumber\\
&&\qquad
+\frac{2}{5}\hat g^2\hat A^{i_1i_2}\hat A^{i_3j}\eta_{jk}\hat A^{ki_4}
\hat A^{i_5l}\eta_{lm}\hat A^{mi_6}\Bigr).
\label{eq:d5n8lag}
\end{eqnarray}
This corresponds to a truncation of maximally symmetric $SO(5,1)$
gauged de Sitter supergravity.  Note that the kinetic terms for the
gauge fields in the compact subgroup $SO(5)\subset SO(5,1)$
have the wrong sign, as expected in the $*$ theory.  In addition,
the potential
\begin{equation}
V=-\frac{1}{2}\hat g^2\Bigr(2\hat T_{ij}\eta^{jk}\hat T_{kl}\eta^{li}
-(\eta^{ij}\hat T_{ij})^2\Bigr),
\end{equation}
has opposite sign from the sphere case, and yields a maximally symmetric
dS$_5$ vacuum, invariant under the de Sitter supergroup $SU^*(4|4)$.
Note that this is the opposite sign of the $SO(6)$ but {\it not} the
$SO(5,1)$ potential of \cite{Gunaydin:1984qu,Pernici:ju,Gunaydin} since in
the present case $T_{ij}\eta^{jk}$ has all positive eigenvalues.

\subsection{Truncation to $D=5$, $N=2$ de Sitter supergravity}

The maximal $N=8$ $SO(6)$ gauged anti-de Sitter supergravity has a
natural $U(1)^3$ truncation to $N=2$ gauged supergravity coupled to two
vector multiplets.  The bosonic fields of this truncation comprise the
metric, two scalars and three gauge fields.  The three gauge fields are
naturally taken to be the mutually commuting subset $A^{12}$, $A^{34}$ and
$A^{56}$ of the full $SO(6)$ gauge fields.  At the same time, the two
scalars originate from the parametrization of $T_{ij}$ as
$T={\rm diag}(X_1,X_1,X_2,X_2,X_3,X_3)$, with the constraint $X_1X_2X_3=1$.

Turning to the de Sitter supergravity, on the other hand, the
maximal compact subgroup $SO(5)\subset SO(5,1)$ does not admit a $U(1)^3$
truncation.  Nevertheless, we may perform an analogous truncation to
two compact and one non-compact $U(1)$ gauge fields.  To do so, we let
\begin{equation}
A^1=\hat A^{12},\qquad
A^2=\hat A^{34},\qquad
A^3=\hat A^{56},
\end{equation}
where $A^3$ is the non-compact gauge field.  In addition, the
scalars may be given by $\hat T={\rm diag}(X_1,X_1,X_2,
X_2,X_3,-X_3)$, with $X_1X_2X_3=1$.

Since the choice of $U(1)^3$ truncation corresponds to taking mutually
commuting rotation (boost) planes along the 1-2, 3-4 and 5-6 directions,
it is natural to parametrize the hyperboloid $H^5$ by taking
\begin{equation}
\hat \mu=\{\mu_1\sin\phi_1,\mu_1\cos\phi_1,\mu_2\sin\phi_2,
\mu_2\cos\phi_2,\mu_3\sinh\phi_3,\mu_3\cosh\phi_3\}.
\end{equation}
In this case, the constraint $\eta_{ij}\hat\mu^i\hat\mu^j=-1$ turns into
$\mu_1^2+\mu_2^2-\mu_3^2=-1$, and the reduction ansatz
(\ref{eq:b5redans}) becomes
\begin{eqnarray}
ds_{10}^2&=&\Delta^{1/2}ds_5^2+\hat g^{-2}\Delta^{-1/2}\sum_{i=1}^3
X_i^{-1}\bigl(\eta_i d\mu_i^2+\mu_i^2(d\phi_i+\hat g A^i)^2),\nonumber\\
\hat F_{(5)}&=&\hat G_{(5)}+*\hat G_{(5)},\nonumber\\
\hat G_{(5)}&=&2\hat g\sum_{i=1}^3(\eta_iX_i^2\mu_i^2+\Delta
X_i)\epsilon_{(5)}+\fft12\hat g^{-1}\sum_{i=1}^3 \eta_i X_i^{-1}\bar *
dX_i\wedge d(\mu_i^2)\nonumber\\
&&+\fft12\hat g^{-2}\sum_{i=1}^3 \eta_iX_i^{-2}d(\mu_i^2)\wedge
(d\phi_i-\hat gA^i)\wedge\bar *F^i,
\end{eqnarray}
where $\Delta=-\sum_{i=1}^3\eta_iX_i\mu_i^2$ and $\eta_i=(+1,+1,-1)$,
signifying the non-compact nature of $H^5$.
Inserting the above $U(1)^3$ truncation into (\ref{eq:d5n8lag}), we
obtain the Lagrangian describing the bosonic sector of the
$N=2$ theory:
\begin{equation}
e^{-1}{\cal L}_5 = R\bar *1-\ft12\bar*d\varphi_1\wedge d\varphi_1
-\ft12\bar*d\varphi_2\wedge d\varphi_2-V\bar*1+\ft12\sum_{i=1}^3
\eta_iX_i^{-2}\bar*F^i\wedge F^i-F^1\wedge F^2\wedge A^3.
\label{eq:d5n2lag}
\end{equation}
Here we have chosen to parametrize the scalars in terms of two dilatons
$\vec\varphi=\{\varphi_1,\varphi_2\}$ according to
\begin{equation}
X_i=e^{-\fft12\vec a_i\cdot\vec\varphi},\qquad
\vec a_i\cdot\vec a_j=4\delta_{ij}-\ft43.
\end{equation}

Note that the first two compact gauge fields in
(\ref{eq:d5n2lag}) have wrong sign kinetic terms, while the non-compact
gauge field, which would ordinarily have come in with the wrong sign,
now enters with the proper one.  The potential is positive definite,
and has the form $V=4\hat g^2 \sum_i{X_i^{-1}}$.  While we have
started with a truncation of the $D=5$, $N=8$ theory, the $N=2$ content
is complete.  Thus from an $N=2$ perspective, we have obtained a
consistent reduction of IIB$^*$ theory (followed by truncation) to
$N=2$ de Sitter supergravity in five dimensions coupled to two vector
multiplets, at least in the bosonic sector.

We may further truncate the bosonic Lagrangian (\ref{eq:d5n2lag}) by
setting $F^1=F^2=F/\sqrt{2}$ along with
$X_1=X_2=X_3^{-1/2}=e^{-\fft1{\sqrt{6}}\varphi}$.  The resulting
Lagrangian is that of $N=2$ de Sitter supergravity coupled to a single
vector multiplet
\begin{equation}
e^{-1}{\cal L}_5 = R\bar *1-\ft12\bar*d\varphi\wedge d\varphi
-4\hat g^2(2e^{\fft1{\sqrt{6}}\varphi}+e^{-\fft2{\sqrt{6}}\varphi})\bar*1
+\ft12e^{\fft2{\sqrt{6}}\varphi}\bar*F\wedge
F-\ft12e^{-\fft4{\sqrt{6}}\varphi}
\bar*{\cal F}\wedge{\cal F}-\ft12F\wedge F\wedge{\cal A}.
\label{eq:d5n2lag2}
\end{equation}
Here ${\cal A}=A^3$ denotes the non-compact gauge field.

The anti-de Sitter supergravity, where both $F$ and ${\cal F}$ have
proper kinetic terms, admits one further truncation to eliminate the
remaining vector multiplet by setting $\varphi=0$ and $A=\sqrt{2}{\cal
A}$.  However, in this case, a consistent truncation to $\varphi=0$
involves satisfying the condition $\bar*F\wedge F+\bar*{\cal F}\wedge
{\cal F}=0$, which arises from the $\varphi$ equation of motion.  Since
this condition cannot be met for real gauge fields, we are unable to
reduce the de Sitter theory of (\ref{eq:d5n2lag2}) any further.

It should be noted, of course, that were one to simply analytically
continue from the truncated $N=2$ anti-de Sitter supergravity Lagrangian
to obtain the pure supergravity truncation of (\ref{eq:d5n2lag}), one
would have to set $iF^1=iF^2=F^3=F/\sqrt{3}$ in order to obtain 
\begin{equation}
e^{-1}{\cal L}_5=R\bar*1-12\hat g^2\bar*1
-\ft12\bar*F\wedge F+\ft1{3\sqrt{3}}F\wedge F\wedge A.
\end{equation}
While this appears to yield a desirable theory with proper sign kinetic
term and positive cosmological constant, we emphasize that this cannot
be viewed as a reduction of IIB nor of IIB$^*$ theory, since this would
necessarily involve imaginary (or in general complex) bosonic fields.

\section{Reductions of $M^*$-theory}

In the previous section we have examined the $H^5$ reduction of IIB$^*$
supergravity, which gives rise to a de Sitter supergravity in five
dimensions.  Of course, similar techniques may be applied for the reduction
of $M^*$ theory.  In particular, we now turn to the dS$_4\times$AdS$_7$
reduction of $M^*_{(9,2)}$ theory.  Since this theory has two time
directions, an interesting feature arises in that we have the freedom of
regarding either dS$_4$ or AdS$_7$ as the spacetime, with the other factor
considered as an internal space.  We now examine both possibilities
in detail.

\subsection{The AdS$_7$ reduction of M$^*$ theory}

By taking AdS$_7$ as a compactification space, the resulting reduction
will yield a four-dimensional de Sitter supergravity theory with
non-compact gauge symmetry $SO(6,2)$.  As before, we begin with the 
truncated $S^7$ reduction ansatz of $M$-theory, given by
\cite{Cvetic:1999xx}
\begin{eqnarray}
ds^2_{11}&=&\Delta^{2/3}ds_4^2+g^{-2}\Delta^{-1/3}T_{ij}^{-1}D\mu^iD\mu^j,\\
F_{(4)}&=&-gU\epsilon_{(4)}+g^{-1}((T^{-1})^{ij}\bar{\ast}DT_{jk})
\wedge(\mu^kD\mu^i) 
-\frac{1}{2}g^{-2}(T^{-1})^{ik}(T^{-1})^{jl}\bar{\ast} F^{ij}_{(2)}
\wedge D\mu^k \wedge D\mu^l,\nonumber\\
\end{eqnarray}
where
\begin{equation}
U=2T_{ij}T_{jk} \mu^i \mu^k -\Delta T_{ii},\qquad\Delta=T_{ij}\mu^i \mu^j,
\end{equation}
and $\delta_{ij}\mu^i\mu^j=1$ (with $i,j=1,2,\ldots,8$) so that the
$\mu^i$ coordinates parametrize a seven-sphere.  This ansatz
retains the full set of $SO(8)$ Yang-Mills fields, $A_{(1)}^{[ij]}$,
which satisfy the standard relations given by (\ref{eq:Ddef}).  In
addition, there are 35 scalars represented by the symmetric unimodular
matrix $T_{ij}$, and which are described by the coset $SL(8,R)/SO(8)$.
Note that the 35 pseudo-scalars of $D=4$, $N=8$ have been truncated
away, and as a result this is technically not a consistent
truncation.  Nevertheless, this ansatz may be used to lift a large
class of four-dimensional solutions without axions.

The M$^*_{(9,2)}$ supergravity may be formally obtained from M-theory by
analytically continuing the four-form, $F_{(4)}\to i\hat F_{(4)}$, while
simultaneously performing a Wick rotation on one of the spatial
coordinates (so as to yield a theory with two times).  For the
Freund-Rubin ansatz leading to AdS$_4\times S^7$, it is clear that the
Wick rotation should be performed on one of the seven sphere
coordinates, thus yielding AdS$_4\times$dS$_7$.  The analytic
continuation on $F_{(4)}$ then finally gives the dS$_4\times$AdS$_7$
solution of M$^*_{(9,2)}$.  As a result, this leads us to make the
continuation $g\to i\hat g$ along with a reparametrization of the sphere
coordinates
\begin{equation}
\mu^i\to i\hat \mu^i,\qquad \mu^m\to\hat\mu^m\qquad
(i=1,\ldots,6,\quad m=7,8).
\end{equation}
The resulting $\hat\mu$'s now parametrize AdS$_7$ in terms of an
$SO(6,2)/SO(6,1)$ coset.

In addition, we must analytically continue the gauge fields
\begin{equation}
A_{(1)}^{ij}\to -i\hat A_{(1)}^{ij},\qquad
A_{(1)}^{[im]}\to -\hat A_{(1)}^{[im]},\qquad
A_{(1)}^{[mn]}\to i\hat A_{(1)}^{[mn]}\qquad
(i,j=1,\ldots,6,\quad m,n=7,8),
\end{equation}
as well as the scalar matrix
\begin{equation}
T_{ij}\to \hat T_{ij},\qquad
T_{(im)}\to i\hat T_{(im)},\qquad
T_{mn}\to-\hat T_{mn}\qquad (i,j=1,\ldots,6, \quad m,n=7,8).
\end{equation}
The analytic continuation of the gauge fields leads to a non-compact
gauge group $SO(6,2)$.  As a result, the reduction ansatz for
M$^*_{(9,2)}$ theory on dS$_4\times AdS_7$ is given by
\begin{eqnarray}
ds^2_{11}&=&\Delta^{2/3}ds_4^2+\hat{g}^{-2}\Delta^{-1/3}
\eta_{ij}(\hat T^{-1})^{jk}\eta_{kl}D\hat\mu^{i}D\hat\mu^{l},\nonumber\\
\hat{F}_{(4)}&=&-\hat{g}U\epsilon_{(4)}
+\hat{g}^{-1}(\eta_{ij}(\hat T^{-1})^{jk}\bar{\ast}D\hat T_{kl})\wedge
(\hat{\mu}^{l}D\hat\mu^{i})\nonumber\\
&&\qquad-\frac{1}{2}\hat{g}^{-2}\eta_{ij}(\hat T^{-1})^{jk}\eta_{kl}
\eta_{mn}(\hat T^{-1})^{np}\eta_{pq}\bar{\ast}
\hat{F}^{im}_{(2)} \wedge (D\hat\mu^{l} \wedge D\hat\mu^{q}),
\label{eq:b4redans}
\end{eqnarray}
where
\begin{equation}
U=-2\hat T_{ij}\eta^{jk}\hat T_{kl} \hat{\mu}^i \hat{\mu}^l
-\Delta \eta^{ij}\hat T_{ij},\qquad
\Delta=-\hat T_{ij} \hat{\mu}^i \hat{\mu}^j,\qquad
\eta_{ij}\hat{\mu}^i \hat{\mu}^j = -1.
\end{equation}
Here, $\eta_{ij}={\rm diag}(+,+,+,+,+,+,-,-)$ is the $SO(6,2)$ invariant
metric.  The $SO(6,2)$ covariant derivatives may be written in a
straightforward manner using $\eta_{ij}$ when appropriate, and have the
same structure as those of (\ref{eq:nccovar}).

The resulting four-dimensional Lagrangian has the form
\begin{eqnarray}
{\cal L}_4&=&R\,\bar{\ast}1-\frac{1}{4}(\hat T^{-1})^{ij}\bar{\ast}D\hat
T_{jk}\wedge (\hat T^{-1})^{kl}D\hat T_{li}
+\frac{1}{4}\eta_{ij}(\hat T^{-1})^{jk}\eta_{kl}
\eta_{mn}(\hat T^{-1})^{np}\eta_{pq}
\bar{\ast}\hat F^{im}_{(2)}\wedge\hat F^{lq}_{(2)}\nonumber\\
&&\qquad-V\,\bar{\ast}1,
\label{eq:d4n8lag}
\end{eqnarray}
and corresponds to a truncation (without pseudoscalars) of the bosonic
sector of $N=8$, $SO(6,2)$ gauged de Sitter supergravity.  The potential
is given by
\begin{equation}
V=-\frac{1}{2}\hat g^2\Bigr(2\hat T_{ij}\eta^{jk}\hat T_{kl}\eta^{li}
-(\eta^{ij}\hat T_{ij})^2\Bigr),
\end{equation}
and yields a maximally symmetric dS$_4$ vacuum.  In addition, the gauge
fields in the compact subgroup $SO(6)\times SO(2) \subset SO(6,2)$ have
wrong sign kinetic terms.

\subsection{Truncation to $D=4$, $N=2$ de Sitter supergravity}

Although the $N=8$ de Sitter supergravity has a non-compact $SO(6,2)$
gauge group, it nevertheless admits a natural $U(1)^4 \subset SO(6)\times SO(2)
\subset SO(6,2)$ truncation to $N=2$ de Sitter supergravity coupled to
three vector multiplets.  The four gauge fields are naturally taken as
\begin{equation}
A^1=\hat A^{12},\qquad
A^2=\hat A^{34},\qquad
A^3=\hat A^{56},\qquad
A^4=\hat A^{78},
\end{equation}
where we follow the notation of the previous section that
$i,j=1,\ldots,6$ are $SO(6)$ indices, while $m,n=7,8$ are $SO(2)$
indices.  This choice of parametrization also suggests that we take
\begin{equation}
\hat \mu=\{\mu_a\sin\phi_a,\mu_a\cos\phi_a\},\qquad a=1,\ldots,4.
\end{equation}
Note that $\phi_4$ has the conventional interpretation as a periodic
AdS$_7$ time coordinate.  Thus $A^4$, the $SO(2)$ gauge field, is
necessarily connected to gauging the isometry related to the second time
direction of the $M^*_{(9,2)}$ theory.

Corresponding to this choice of truncation, we take the surviving
scalars to be given by $\hat T={\rm diag}(X_1,X_1,X_2,
X_2,X_3,X_3,-X_4,-X_4)$, where $X_1X_2X_3X_4=1$.  The background AdS$_7$
geometry is determined by the constraint $\eta_{ij}\hat\mu^i\hat\mu^j=-1$,
or equivalently $\mu_1^2+\mu_2^2+\mu_3^2-\mu_4^2=-1$.  The truncation of
(\ref{eq:b4redans}) is then
\begin{eqnarray}
ds_{11}^2&=&\Delta^{2/3}ds_4^2+\hat g^{-2}\Delta^{-1/3}\sum_{i=1}^4
X_i^{-1}\eta_i\bigl(d\mu_i^2+\mu_i^2(d\phi_i+\hat g A^i)^2),\nonumber\\
\hat F_{(4)}&=&2\hat g\sum_{i=1}^4(\eta_iX_i^2\mu_i^2+\Delta
X_i)\epsilon_{(4)}+\fft12\hat g^{-1}\sum_{i=1}^4 \eta_i X_i^{-1}\bar *
dX_i\wedge d(\mu_i^2)\nonumber\\
&&+\fft12\hat g^{-2}\sum_{i=1}^4 \eta_iX_i^{-2}d(\mu_i^2)\wedge
(d\phi_i-\hat gA^i)\wedge\bar *F^i,
\end{eqnarray}
where $\Delta=-\sum_{i=1}^4\eta_iX_i\mu_i^2$ and $\eta_i=(+1,+1,+1,-1)$.
The resulting theory may be described by the Lagrangian
\begin{equation}
e^{-1}{\cal L}_4 = R\bar *1-\ft12\sum_{\alpha=1}^3\bar*d\varphi_\alpha\wedge
d\varphi_\alpha+\ft12\sum_{i=1}^4 X_i^{-2}\bar*F^i\wedge F^i-V\bar*1,
\label{eq:d4n2lag}
\end{equation}
where
\begin{equation}
V=4\hat g^2 \sum_{i<j}{X_iX_j}=8\hat g^2\sum_{\alpha=1}^3\cosh\varphi_\alpha
\end{equation}
is positive definite.  Here the $X_i$ scalars are parametrized in terms of
three dilatons $\vec\varphi=\{\varphi_1,\varphi_2,\varphi_3\}$ according to
\begin{equation}
X_i=e^{-\fft12\vec a_i\cdot\vec\varphi},\qquad
\vec a_i\cdot\vec a_j=4\delta_{ij}-1.
\end{equation}
This Lagrangian is simply the $U(1)^4$ truncation of (\ref{eq:d4n8lag}).
Unlike the $D=5$, $N=2$ de Sitter theory of (\ref{eq:d5n2lag2}), here
all four $U(1)$ fields are compact (and all have wrong sign kinetic
terms).  Note furthermore that, as already mentioned previously,
(\ref{eq:d4n2lag}) cannot be viewed as a consistent $N=2$ reduction of
$M^*_{(9,2)}$ theory, as it is missing three axionic scalars.  Hence
the vector multiples are incomplete.  This truncation with only dilatonic
scalars is similar to that considered in \cite{Duff:1999gh} for the
truncated $N=8$ anti-de Sitter supergravity.

We may further truncate the bosonic Lagrangian (\ref{eq:d4n2lag}) by
setting $F^1=F^2=F/\sqrt{2}$ and $F^3=F^4={\cal F}/\sqrt{2}$ along with
$X_1=X_2=X_3^{-1}=X_4^{-1}=e^{\fft12\varphi}$.  The resulting
Lagrangian is that of $N=2$ de Sitter supergravity coupled to a single
vector multiplet
\begin{equation}
e^{-1}{\cal L}_4 = R\bar *1-\ft12\bar*d\varphi\wedge d\varphi
-8\hat g^2(2+\cosh\varphi)\bar*1
+\ft12e^{-\varphi}\bar*F\wedge F +\ft12e^{\varphi}\bar*{\cal F}\wedge
{\cal F}.
\label{eq:d4n2lag2}
\end{equation}
While this theory is still incomplete in that it lacks an axion, one
further truncation to pure $N=2$ de Sitter supergravity is possible by
equating the $U(1)$ fields and setting $\varphi=0$.  The pure
supergravity Lagrangian is
\begin{equation}
e^{-1}{\cal L}_4 = R\bar *1 -24\hat g^2\bar*1 +\ft12\bar*F\wedge F,
\label{eq:d4n2lag3}
\end{equation}
and concisely captures both features of * theory, namely a positive
cosmological constant and wrong-sign kinetic term for the graviphoton.

\subsection{The dS$_4$ reduction of M$^*$ theory}

We now consider the second possibility for interpreting the
dS$_4\times$AdS$_7$ compactification of $M^*_{(9,2)}$ theory, namely
where we view the resulting theory as a seven dimensional anti-de
Sitter supergravity with $SO(4,1)$ gauging.  This case is especially
interesting in that the {\it complete} consistent Kaluza-Klein
reduction of ordinary eleven-dimensional supergravity to seven
dimensions is known from the work of \cite{Nastase:1999cb,Nastase}.
Thus, unlike in the cases considered above, there is no need to
truncate.

To obtain the dS$_4$ reduction of $M^*_{(9,2)}$ theory, we analytically
continue and Wick rotate the ansatz obtained in \cite{Nastase,Cvetic1},
which is given by
\begin{eqnarray}
ds_{11}&=&\Delta^{1/3}ds_7^2+g^{-2}\Delta^{-2/3}T_{ij}^{-1}D\mu^iD\mu^j,
\nonumber\\
\ast_{11} F_{(4)}&=&-gU\epsilon_{(7)}-g^{-1}(T_{ij}^{-1}\bar{\ast} DT_{jk})
\wedge(\mu^kD\mu^i)  
+ \frac{1}{2}g^{-2}T_{ik}^{-1}T_{jl}^{-1}\bar{\ast} F^{ij}_{(2)}\wedge
D\mu^k \wedge D\mu^l\nonumber\\
&&+ g^{-4}\Delta^{-1}T_{ij}S_{(3)}^i\mu^j\wedge W 
-\frac{1}{6}g^{-3}\Delta^{-1}\epsilon_{ijl_1l_2l_3}\bar{\ast}S_{(3)}^m
T_{im}T_{jk}\mu^k \wedge D\mu^{l_1}\wedge D\mu^{l_2}\wedge D\mu^{l_3},
\nonumber\\
\end{eqnarray}
where
\begin{eqnarray}
&U=2T_{ij}T_{jk} \mu^i \mu^k -\Delta T_{ii},\qquad
\Delta=T_{ij}\mu^i \mu^j,&\nonumber\\
&W=\ft{1}{4!}\epsilon_{i_1i_2i_3i_4i_5}\mu^{i_1}D\mu^{i_2}\wedge D\mu^{i_3}
\wedge D\mu^{i_4}\wedge D\mu^{i_5}.&
\end{eqnarray}
The full reduction gives rise to $N=4$ gauged $SO(5)$ supergravity in
seven dimensions.  The bosonic fields consist of 14 scalars given by the
symmetric unimodular $T_{ij}$ which describe the coset $SL(5,R)/SO(5)$,
the $SO(5)$ gauge fields $A^{[ij]}_{(1)}$, and 5 three-form potentials
$S^i_{(3)}$ transforming in the fundamental of $SO(5)$.
Note that the reduction is given on the dual of the eleven-dimensional
four-form, $\ast_{11} F_{(4)}$, with $\epsilon_{(7)}$ corresponding to the
volume 7-form in spacetime.  In addition, the coordinates $\mu^i$,
subject to the constraint $\delta_{ij}\mu^i\mu^j=1$, parameterize the
unit four-sphere.

The analytic continuation to * theory is different from the previous
cases, since here this is no need to take $g\to i\hat g$.  This is
because the simultaneous continuation $F_{(4)}\to i\hat F_{(4)}$ along
with Wick rotation of one of the $S^4$ coordinates leaves the present
four-form ansatz unchanged.  To see this, consider the
pure AdS$_7\times S^4$ solution, given essentially by $F_{(4)}=
g\, dx^7\wedge dx^8\wedge dx^9\wedge dx^{10}$.  While $F_{(4)}$ picks up
an $i$, so does, say, $dx^{10}$.  The only important modification is
then to replace the $S^4$ coset structure $SO(5)/SO(4)$ by the dS$_4$
coset $SO(4,1)/SO(3,1)$ by analytic continuation on the $\mu_i$'s.
The combined transformation, including that of 1-form and 3-form potentials
is given by
\begin{eqnarray}
&&\kern-2em
\mu^i\to \hat \mu^i,\qquad \mu^5\to i\hat\mu^5,\nonumber\\
&&\kern-2em
S_{(3)}^i\to \hat S_{(3)}^i, \qquad S_{(3)}^5\to i\hat S_{(3)}^5,\qquad
A_{(1)}^{[ij]}\to \hat A_{(1)}^{[ij]},\qquad
A_{(1)}^{[i5]}\to i\hat A_{(1)}^{[i5]}\qquad (i,j=1,\ldots,4).\qquad
\end{eqnarray}
In order for $U$ and $\Delta$ to be real, we also continue
appropriate entries in the $T_{ij}$ matrix, namely
\begin{equation}
T_{ij}\to \hat T_{ij},\qquad
T_{(i5)}\to -i\hat T_{(i5)} ,
\qquad T_{55}\to-\hat T_{55}\qquad (i,j=1, \ldots,4).
\end{equation}
The resulting reduction ansatz for $M^*_{(9,2)}$ theory on
dS$_4\times$AdS$_7$ may be given in terms of the $SO(4,1)$ metric
$\eta_{ij}={\rm diag}(+,+,+,+,-)$.  We find
\begin{eqnarray}
ds^2_{11}&=&\Delta^{1/3}ds_7^2+{g}^{-2}\Delta^{-2/3}
\eta_{ij}(\hat T^{-1})^{jk}\eta_{kl}D\hat\mu^{i}D\hat\mu^{l},\nonumber\\
\ast_{11}\hat{F}_{(4)}&=&-{g}U\epsilon_{(7)}
-{g}^{-1}(\eta_{ij}(\hat T^{-1})^{jk}\bar{\ast}D\hat T_{kl})\wedge
(\hat{\mu}^{l}D\hat\mu^{i})\nonumber\\
&&\qquad+\frac{1}{2}g^{-2}\eta_{ij}(\hat T^{-1})^{jk}\eta_{kl}
\eta_{mn}(\hat T^{-1})^{np}\eta_{pq}\bar{\ast}
\hat{F}^{im}_{(2)} \wedge D\hat\mu^{l} \wedge D\hat\mu^{q}\nonumber\\
&&\qquad
+g^{-4}\Delta^{-1}\hat T_{ij}\hat S_{(3)}^i\hat \mu^j\wedge\hat W\nonumber\\
&&\qquad
-\frac{1}{6}g^{-3}\Delta^{-1}\hat\epsilon_{ijl_1l_2l_3}\eta^{in}\eta^{jp}
\bar\ast\hat S_{(3)}^m \hat T_{nm}\hat T_{pk}\mu^k
\wedge \hat D\mu^{l_1}\wedge \hat D\mu^{l_2} \wedge \hat D\mu^{l_3},
\label{eq:b7redans}
\end{eqnarray}
where
\begin{eqnarray}
&U=2\hat T_{ij}\eta^{jk}\hat T_{kl} \hat{\mu}^i \hat{\mu}^l
-\Delta \eta^{ij}\hat T_{ij},\qquad
\Delta=\hat T_{ij} \hat{\mu}^i \hat{\mu}^j,\qquad
\eta_{ij}\hat{\mu}^i \hat{\mu}^j = 1,&\nonumber\\
&\hat W=\ft{1}{4!}\hat\epsilon_{i_1i_2i_3i_4i_5}\hat \mu^{i_1}
\hat D\mu^{i_2}\wedge\hat D\mu^{i_3}\wedge \hat D\mu^{i_4}\wedge\hat
D\mu^{i_5}.&
\end{eqnarray}
Note that $\hat\epsilon_{ijklm}$ is the $SO(4,1)$ invariant
antisymmetric tensor.

The resulting Lagrangian has the form
\begin{eqnarray}
{\cal L}_7&=&R\,\bar{\ast}1-\frac{1}{4}(\hat T^{-1})^{ij}\bar{\ast}D\hat
T_{jk}\wedge (\hat T^{-1})^{kl}D\hat T_{li}
-\frac{1}{4}\eta_{ij}(\hat T^{-1})^{jk}\eta_{kl}
\eta_{mn}(\hat T^{-1})^{np}\eta_{pq}
\bar{\ast}\hat F^{im}_{(2)}\wedge\hat F^{lq}_{(2)} \nonumber\\
&&\qquad-\frac{1}{2}\hat T_{ij}\bar\ast\hat S_{(3)}^i\wedge\hat S_{(3)}^j 
+\frac{1}{2g}\eta_{ij}\hat S_{(3)}^i\wedge \hat H_{(4)}^j 
-\frac{1}{8}\hat\epsilon_{ij_1j_2j_3j_4}\hat S_{(3)}^i\wedge
\hat F_{(2)}^{j_1j_2}
\wedge \hat F_{(2)}^{j_3j_4} -V\,\bar{\ast}1\nonumber\\
&&\qquad+\hbox{(Chern-Simons)},
\label{eq:d7n4lag}
\end{eqnarray}
where 
\begin{equation}
\hat H_{(4)}^i=d\hat S_{(3)}^i+g\hat A^{ik}\eta_{kj}\hat S_{(3)}^j.
\end{equation}
This resulting theory is simply an unusual non-compact gauging of maximal
$D=7$ supergravity.  In particular, both the $SO(4,1)$ gauge fields and the
potential
\begin{equation}
V=\frac{1}{2}g^2\Bigr(2\hat T_{ij}\eta^{jk}\hat T_{kl}\eta^{li}
-(\eta^{ij}\hat T_{ij})^2\Bigr),
\end{equation}
have their usual signs, despite the $M^*_{(9,2)}$ origin of this theory.

\subsection{Truncation of the $D=7$ theory}

The $D=7$, $N=4$ theory constructed above admits a truncation to $N=2$
supergravity coupled to an $N=2$ vector.  The fields in the supergravity
multiplet consist of the metric $g_{\mu\nu}$, three gauge bosons in the
adjoint of $SU(2)_+\subset SO(4)\subset SO(4,1)$, namely
$\hat A_{(1)}^{\overline{ij}}+\fft12\epsilon^{\overline{ijkl}}
\hat A_{(1)}^{\overline{kl}}$, a single three-form potential $\hat
S_{(3)}^5$, and a dilaton scalar $\varphi$.  In addition, the vector
multiplet consists of a single $U(1)$ gauge potential contained in
$SO(2)_-$ and three axionic scalars.
In principle, it is straightforward to consistently truncate the above
reduction to yield the $N=2$ theory.  However, the presence of the
non-abelian graviphotons and the axionic scalars results in some
complication.  We choose here not to pursue this truncation, but instead
focus on a two $U(1)$ truncation with $U(1)^2\subset SO(4,1)$, while
simultaneously retaining two dilatonic scalars \cite{Cvetic}.  While
this is not a consistent truncation, it may still be used to lift
solutions to eleven dimensions.

To obtain this $U(1)^2$ truncation of the full theory, we parametrize
the internal $dS_4$ by $(\mu_1\sin{\phi_1},\mu_1\cos{\phi_1},
\mu_2\sin{\phi_2},\mu_2\cos{\phi_2},\mu_0)$.  Only two gauge fields,
$A^1=\hat A_{(1)}^{12}$ and $A^2=\hat A_{(1)}^{34}$, are kept as well as 
$\hat T={\rm diag}(X^1, X^1, X^2, X^2, -X^0)$.  The rest of the fields are
truncated.  The reduction ansatz, (\ref{eq:b7redans}), then becomes 
\begin{eqnarray}
ds_{11}^2=\Delta^{1/3}ds_7^2+g^{-2}\Delta^{-2/3}\bigl(-X_0^{-1}d\mu_0^2+
\sum_{i=1}^2
X_i^{-1}\bigl(d\mu_i^2+\mu_i^2(d\phi_i+g A^i)^2),\nonumber\\
\ast_{11} \hat F_{(4)}= 2 g\sum_{i=1}^2(X_i^2\mu_i^2-\Delta
X_i)\epsilon_{(7)}-g(2X_0^2\mu_0^2-\Delta X_0)\epsilon_{(7)} 
+\fft12 g^{-1}\sum_{i=1}^2 X_i^{-1}\bar *dX_i\wedge d(\mu_i^2)\nonumber\\
-\fft12 g^{-1}X_0^{-1}\bar\ast dX_0\wedge d(\mu_0^2)
+\fft12\hat g^{-2}\sum_{i=1}^2 \eta_iX_i^{-2}d(\mu_i^2)\wedge
(d\phi_i-\hat gA^i)\wedge\bar *F^i,\quad
\end{eqnarray}
where $\Delta=-X_0\mu_0^2+\sum_{i=1}^2X_i\mu_i^2$.  
Inserting the above $U(1)^2$ truncation into (\ref{eq:d7n4lag}), we
obtain the bosonic Lagrangian
\begin{equation}
e^{-1}{\cal L}_7 = R\bar *1-\ft12\bar*d\varphi_1\wedge d\varphi_1
-\ft12\bar*d\varphi_2\wedge d\varphi_2 
-\ft12\sum_{i=1}^2 X_i^{-2}\bar*F^i\wedge F^i-V\bar*1.
\label{eq:d7n2lag}
\end{equation}
Here we have chosen to parametrize the scalars in terms of two dilatons
$\vec\varphi=\{\varphi_1,\varphi_2\}$ according to
\begin{equation}
X_i=e^{-\fft12\vec a_i\cdot\vec\varphi},\qquad
\vec a_i\cdot\vec a_j=4\delta_{ij}-\fft{8}{5},\qquad
X_0=(X_1X_2)^{-2} \qquad(i,j=1,2).
\end{equation}
The potential is given by 
\begin{equation}
V = -g^2(4X_1X_2+2X_0X_1+2X_0X_2-\fft12 X_0^2).
\end{equation}
corresponding to the AdS$_7$ vacuum.  Further truncation is possible by
setting $\varphi_1=0$ and $F^1=F^2=F/\sqrt{2}$. In this case, the
resulting Lagrangian becomes 
\begin{equation}
e^{-1}{\cal L}_7 = R\bar *1 -\ft12 \bar*\partial\varphi\wedge\partial\varphi
-\ft12 X^{-2}\bar *F\wedge F+g^2(4X^2+4X^{-3}-\ft12 X^{-8})\bar*1.
\label{eq:d7n2lag2}
\end{equation}
Note that both truncations (\ref{eq:d7n2lag}) and (\ref{eq:d7n2lag2})
have gauge fields contained in the maximally compact subgroup
$SO(4)\subset SO(4,1)$.  As a result, they both have proper kinetic
energy terms and a standard potential admitting the AdS vacuum.  So,
after appropriate truncation, we have in fact obtained a unitary theory
(indistinguishable from the corresponding truncation of ordinary
$N=4$ gauged supergravity in seven dimensions) from a non-unitary one.
Nevertheless, this fact is perhaps of limited usefulness, as the
model before truncation inherits the usual drawbacks of the
underlying field theory description of * theory.

\section{Lifting de Sitter black hole solutions}

It is an interesting observation that, even in the absence of obvious
supersymmetry, multi-centered de Sitter black hole solutions are known to
exist, and have been constructed in \cite{Kastor:1992nn,London} in the
context of Einstein-Maxwell theory with a positive cosmological constant.
These solutions resemble traditional BPS objects in that they satisfy a
set of first order equations \cite{London,Liu,Behrndt} related to the
analytic continuation of the Killing spinor equation in AdS gauged
supergravity.  In the $d$-dimensional Reissner-Nordstrom case (for $d\ge4$),
the solution may be written in terms of a cosmological metric
\begin{equation}
ds^2=-H^{-2}(t,\vec x\,)dt^2+H^{2/(d-3)}(t,\vec x\,)\,a^2(t)d\vec x\,^2,
\label{eq:rnmet}
\end{equation}
where the scale factor is given by $a(t)=e^{\hat gt}$ and the harmonic
function has the multi-center form
\begin{equation}
H(t,\vec x\,)=1+\sum_i\fft{q_i}{(a(t)|\vec x-\vec x_i|)^{d-3}}.
\end{equation}
It is apparent that while the background is time-dependent, this time
dependence is rather trivial and is simply a reflection of the cosmological
expansion in de Sitter space.  For this reason, these black holes are
natural candidates to lift to the higher dimensional * theory, where they
may be interpreted as possible fundamental objects of the * theory.

It was demonstrated in \cite{Liu} that this Reissner-Nordstrom black hole,
(\ref{eq:rnmet}), may be generalized to the analytically continued $D=5$,
$N=2$ supergravity, and in particular the STU model corresponding to the
anti-de Sitter version of (\ref{eq:d5n2lag}).  Further generalizations 
to Einstein-Maxwell-dilaton theories in arbitrary dimensions are also
possible.

\subsection{$D=5$ dS black holes}

We first examine the five-dimensional case obtained by the hyperbolic
reduction of IIB$^*$.  The $U(1)^3$ truncated theory of (\ref{eq:d5n2lag})
admits a three-charge black hole solution, given by
\begin{eqnarray}
ds_5^2&=&-(H_1H_2H_3)^{-2/3}dt^2 + (H_1H_2H_3)^{1/3}e^{2\hat gt}d\vec x\,^2,
\nonumber\\
X_i&=&H_i^{-1}(H_1H_2H_3)^{1/3}, \nonumber\\
A_{(1)}^{1,2}&=&i\left(1-\frac{1}{H_{1,2}}\right)dt,\qquad
A_{(1)}^3 = \left(1-\frac{1}{H_3}\right)dt,
\label{eq:bhfive}
\end{eqnarray}
where
\begin{equation}
H_i(t,\vec x\,)=1 + e^{-2\hat{g}t}\sum_j{\frac{q_i^j}{|\vec x-\vec x_j|^2}}.
\end{equation}
This solution was given implicitly in \cite{Liu}, at least up to analytic
continuation to the present * theory vacuum.  The continuation is chosen
here so that the metric remains real at the expense of introducing
imaginary background gauge fields%
\footnote{The de Sitter black hole solution is a real solution for positive
cosmological constant and correct sign kinetic terms for the fields.
Since the first two $U(1)$'s of (\ref{eq:d5n2lag}) have the wrong sign,
the corresponding gauge potentials become imaginary in (\ref{eq:bhfive}).}.
In fact, the complexification of the solution compensates for the
wrong sign kinetic terms in (\ref{eq:d5n2lag}) in just such a way that the
black hole solution has positive energy compared with the vacuum.

There is a potential difficulty associated with lifting the solution
(\ref{eq:bhfive}) to ten dimensions, in that the imaginary gauge fields
$A^1_{(1)}$
and $A^2_{(1)}$ would lead to a complex ten dimensional metric (as well as
$F_{4}$).  To avoid this, we turn off the first two gauge fields, and lift
the single charge ($A^3_{(1)}$) black hole to IIB$^*$ theory.  The
resulting metric is
\begin{equation}
ds_{10}^2 = \widetilde\Delta^{1/2}(-H^{-1}dt^2+e^{2\hat gt}
d\vec x\,^2)+\hat g^{-2}\widetilde\Delta^{-1/2}ds^2(\widetilde H^5),
\label{eq:bhten}
\end{equation}
where $\widetilde\Delta\equiv H^{2/3}\Delta=1+(1-H)\sinh^2\alpha$ and
\begin{equation}
ds^2(\widetilde H^5)=d\alpha^2+\sinh^2\alpha\,d\Omega_3^2
+(1-H)\sinh^2\alpha\,d\alpha^2+H\cosh^2\alpha(d\psi-\hat gH^{-1}(1-H)dt)^2
\end{equation}
is the metric on the distorted internal hyperbolic space.  Note that we have
chosen an explicit parametrization of the maximally symmetric $H^5$ as
\begin{equation}
ds^2(H^5)=d\alpha^2+\sinh^2\alpha\,d\Omega_3^2+\cosh^2\alpha\,d\psi^2.
\end{equation}
After applying a change of coordinates, $\hat g\tau=e^{\hat gt}$, and a
rearrangement of terms, the lifted solution (\ref{eq:bhten}) may be
written in the form
\begin{equation}
ds_{10}^2 = \widetilde{\cal H}_0^{-1/2}d\vec x\,^2
+\widetilde{\cal H}_0^{1/2}[-\widetilde\Delta
H^{-1}d\tau^2+\tau^2ds^2(\widetilde H^5)],
\label{eq:bhten2}
\end{equation}
where
\begin{equation}
\widetilde{\cal H}_0=\lim_{\tau\to0}\widetilde{\cal H},\qquad
\widetilde{\cal H}=1+\fft1{\hat g^4\tau^4\widetilde\Delta}.
\end{equation}
We have written $\widetilde{\cal H}$ in this form in anticipation of the
brane interpretation of this lifted solution.

In fact, we recall that the IIB$^*$ theory admits an E4-brane solution of
the form \cite{Hull:1998vg}
\begin{equation}
ds_{10}^2={\cal H}^{-1/2}d\vec x\,^2+{\cal H}^{1/2}(-dt^2+dr^2+r^2d\Omega_4^2),
\end{equation}
where
\begin{equation}
{\cal H}=1+\fft1{\hat g^4|t^2-r^2|^2}.
\end{equation}
This may be most readily compared to the lifted single charge black hole
solution of (\ref{eq:bhten2}) by taking the near brane limit $t\to r$ with
$t^2>r^2$.  In this case, we may make a change of variables,
$t=\tau\cosh\beta$, $r=\tau\sinh\beta$, and drop the one in the harmonic
function.  The resulting dS$_5\times H^5$ metric has the form
\begin{equation}
ds_{10}^2={\cal H}_0^{-1/2}d\vec x\,^2+{\cal H}_0^{1/2}[-d\tau^2
+\tau^2(d\beta^2+\sinh^2\beta\,d\Omega_4^2)],
\end{equation}
where ${\cal H}_0=1/\hat g^4\tau^4$.  It should now be apparent that
(\ref{eq:bhten2}) is a generalization of this solution to the case of
non-zero R-charge in five dimensions.  Recalling that the R-charged AdS
black holes may be interpreted as rotating branes \cite{Cvetic}, we see
here that a similar picture holds, albeit with time as a transverse as
opposed to a longitudinal coordinate.

\subsection{$D=4$ dS black holes}

Turning now to four dimensions, we note that the $U(1)^4$ truncation,
(\ref{eq:d4n2lag}), admits a four charge dS black hole solution, given by
\begin{eqnarray}
ds_4^2&=&-(H_1H_2H_3H_4)^{-1/2}dt^2 + (H_1H_2H_3H_4)^{1/2}e^{2\hat gt}
d\vec x\,^2,\nonumber\\
X_i&=&H_i^{-1}(H_1H_2H_3H_4)^{1/4},\nonumber\\
A_{(1)}^{i}&=&i\left(1-\frac{1}{H_i}\right)dt,
\label{eq:bhfour}
\end{eqnarray}
where
\begin{equation}
H_i(t,\vec x\,) = 1 + e^{-\hat{g}t}\sum_j{\frac{q_i^j}{|\vec x-\vec x_j|}}.
\end{equation}
Again we are faced with the difficulty of imaginary gauge potentials.
However, unlike the five-dimensional case, all fields in the $U(1)^4$
truncation have the wrong sign.  Hence there is no truncation (short of
setting all the charges to zero) that makes the solution real.  For this
reason, any lifting of these black holes to the M$^*$ theory would result
in a complex metric, and hence it is unclear what the physical
significance of these solutions is.

Nevertheless, we note that the analytic continuation of the
Reissner-Nordstrom-de Sitter solution is obtained by setting all four
charges equal in (\ref{eq:bhfour}):
\begin{eqnarray}
ds_4^2&=&-H^{-2}dt^2 + H^2e^{2\hat gt}d\vec x\,^2,\nonumber\\
A_{(1)}&=&i\left(1-\frac{1}{H}\right)dt.
\end{eqnarray}
This may be viewed as a solution to the pure $N=2$ de Sitter supergravity
of (\ref{eq:d4n2lag3}) where $A_{(1)}$ is taken as the graviphoton.

\subsection{$D=7$ AdS black holes}

Finally, we note that the $U(1)^2$ truncation of the seven-dimensional theory,
(\ref{eq:d7n2lag}), admits a two charge AdS black hole solution given by 
\begin{eqnarray}
ds_7^2&=&-(H_1H_2)^{-4/5}dt^2 + (H_1H_2)^{1/5}e^{-2igt}d\vec x\,^2,\nonumber\\
X_i&=&H_i^{-1}(H_1H_2)^{2/5}, \nonumber\\
A_{(1)}^{1,2}&=&\left(1-\frac{1}{H_{1,2}}\right)dt,
\label{eq:bhseven}
\end{eqnarray}
where
\begin{equation}
H_i(t,\vec x\,)=1 + e^{4igt}\sum_j{\frac{q_i^j}{|\vec x-\vec x_j|^4}}.
\end{equation}
Although the gauge fields have correct sign kinetic terms, the negative
cosmological constant formally results a complex metric, as noted in
\cite{Liu}.  The further truncated theory of (\ref{eq:d7n2lag2}) also admits
a solution by setting all $H_i=H$ in (\ref{eq:bhseven}):
\begin{eqnarray}
ds_7^2&=&-H^{-8/5}dt^2 + H^{2/5}e^{-2igt}d\vec x\,^2,\nonumber\\
X&=&H^{-1/5}, \nonumber\\
A_{(1)}&=&\left(1-\frac{1}{H}\right)dt.
\end{eqnarray}

\section{Discussion}

So far we have focused only on the reduction of the bosonic sector of *
theories.  Nevertheless, it should be possible to handle the fermions in a
similar manner through analytic continuation.  Unlike the conventional case,
where analytic continuation from an anti-de Sitter to a de Sitter theory would
complexify the fermions and destroy the matching between bosonic and
fermionic degrees of freedom \cite{Pilch}, here the * theories have a twisted
supersymmetry built in (at the expense of wrong sign kinetic terms).  Thus
the resulting fermion sectors should not have any doubling problem.

Although we do not examine the fermions in detail, the structure of the
de Sitter supersymmetry transformations may be derived by an appropriate
continuation of the anti-de Sitter supergravities.  For example, it is well
known that the $S^5$ compactification of IIB supergravity will lead to
maximal $SO(6)$ gauged supergravity in five dimensions
\cite{Gunaydin:1984qu,Pernici:ju,Gunaydin}.  In addition, the
non-compact $SO(p,6-p)$ gauged case has also been investigated in
\cite{Gunaydin}.  Turning to the $H^5$ compactification of IIB$^*$
supergravity, we have seen in section 3 that the appropriate analytic
continuation requires sending $g$ to $i\hat g$.  This results in an
unconventional de Sitter supersymmetry with noncompact SO(5,1) gauge group.

Recall that the $D=5$ ungauged maximal supergravity fields consists of
one graviton, eight gravitini $\psi^a_\mu$, 27 vector fields $A^{[ab]}_\mu$,
48 spin-$1/2$ fields $\lambda^{abc}$ and 42 scalars $\varphi^{abcd}$, 
where $a,b,\ldots$ are $USp(8)$ indices.  The complete scalars parametrize
the noncompact coset space $E_{6(6)}/USp(8)$.  However, for the truncation
considered in section 3, we specialize to the subgroup $SO(5,1)\subset
SL(6,R)\times SL(2,R)\subset E_{6(6)}$, with scalars parametrized by a
15-bein ${V_{AB}}^{cd}$. This 15-bein may be constructed by starting
with a $SL(6,R)$ matrix $S$ and then taking
\begin{eqnarray}
U^{IJ}{}_{KL}&=&2{S_{[K}}^{[I}{S_{L]}}^{J]},\qquad
V^{IJab}=\fft18(\Gamma^{KL})^{ab}{U^{IJ}}_{KL},\nonumber\\
U_{I\alpha}{}^{J\beta}&=&S_I{}^J\delta_\alpha{}^\beta,\qquad
V_{I\alpha}{}^{ab}=\fft1{2\sqrt{2}}(\Gamma_{K\beta})^{ab}U_{I\alpha}{}^{K\beta},
\end{eqnarray}
where the $SO(5,1)$ Dirac matrices satisfy $\{\Gamma_I,\Gamma_J\}=2\eta_{IJ}$
with $\eta_{IJ}={\rm diag}(+,+,+,+,+,-)$.  In addition,
$\Gamma_{I\alpha}\equiv\Gamma_I$ ($\alpha=1$);
$\Gamma_{I\alpha}\equiv i\Gamma_I\Gamma_0$ ($\alpha=2$) where $\Gamma_0$
anticommutes with the first six Dirac matrices.  Here we are following the
notations and conventions of \cite{Gunaydin}.

The scalar kinetic terms $P_\mu{}^{abcd}$ and the composite connection
${Q_{\mu a}}^b$ are defined through
\begin{equation}
\tilde{V}_{cd}{}^{AB}D_{\mu}{V_{AB}}^{ab}=2{Q_{\mu [c}}^{[a}\delta_{d]}^{b]}
+ {P_{\mu}}^{ab}{}_{cd},
\end{equation}
where $\tilde{V}_{cd}{}^{AB}$ is the inverse of $V_{AB}{}^{cd}$.
In addition, the T-tensor is defined as 
\begin{eqnarray}
{T^a}_{bcd}&=&(2V^{IKae}\tilde{V}_{beJK}-V_{J\alpha}{}^{ae}\tilde
V_{be}{}^{I\alpha})\eta^{JL}\tilde{V}_{cdIL}, \\
T_{ab}&=&{T^c}_{abc}.
\end{eqnarray}
Note that, the matrix $S$ is related to $\hat T_{ij}$ introduced in
section 3 by
\begin{equation}
(\hat T^{-1})^{ij}=S_I{}^iS_J{}^j\eta^{IJ}.
\end{equation}
This simply corresponds to the unconventional * supersymmetry with scalars
parameterizing $SL(6,R)/SO(5,1)$.

Besides introducing this noncompact gauging, we also analytically continue
$g\to i\hat g$ and $F_{(2)} \to -i\hat F_{(2)}$ in order to obtain the
Lagrangian (\ref{eq:d5n8lag}).  Therefore the unconventional supersymmetry
transformation rules for the gravitini and spin-$1/2$ fermions are 
\begin{eqnarray}
\delta\psi_{\mu a}&=&D_{\mu}\epsilon_a - \fft{2i}{45}\hat g
T_{ab}\gamma_{\mu}\epsilon^b +\fft{i}6 (\gamma_\mu{}^{\nu\rho}
-4\delta_\mu^\nu\gamma^{\rho})\hat F_{\nu\rho ab}\epsilon^b,\nonumber\\
\delta\lambda_{abc}&=&\sqrt{2}\gamma^{\mu}P_{\mu\, abcd}\epsilon^d
-\fft{i}{\sqrt{2}}\hat g T_{d[abc]}\epsilon^d 
+\fft{3i}{2\sqrt{2}}\gamma^{\mu\nu}\hat F_{\mu\nu [ab}\epsilon_{c]},
\end{eqnarray}
where 
\begin{eqnarray}
\hat F_{\mu\nu ab}&=&\hat F_{\mu\nu IJ}{V^{IJ}}_{ab}=
\ft14\hat F_{\mu\nu}^{ij}\eta_{im}\eta_{jn}S_K{}^mS_L{}^n(\Gamma_{KL})^{ab}
,\nonumber\\
D_{\mu}\epsilon_a&=&\partial_{\mu} \epsilon_a + {Q_{\mu a}}^b\epsilon_b.
\end{eqnarray}
One should be able to obtain these transformations by direct reduction of
the IIB$^*$ transformations, although we have not directly verified this.

The multi-centered dS black holes of the previous section are half-BPS
solutions of this twisted supersymmetry.  However it is not clear if this
is sufficient to demonstrate their stability.  Unlike for BPS objects in
an ordinary supergravity theory, here the wrong-sign kinetic terms allows
the possibility of excitations above the supersymmetric background that
nevertheless have negative energy.  On the other hand, the existence of
multi-centered solutions is at least suggestive that there may be a hidden
symmetry ensuring their stability.

Although the de Sitter supergravities which we have investigated involve
wrong sign kinetic terms and are hence ill-behaved as field theories,
such problems were already present in the underlying * theory.  Thus we
may expect that whatever stringy phenomenon cures the behavior of * theory
would also stabilize the ensuing de Sitter supergravities.  On the other
hand, it is possible that the reduction on non-compact internal
manifolds may yield inherently unstable lower dimensional theories.  Of
course, consistent truncations are possible in a standard AdS$\times$Sphere
reduction, even when states are not well separated by {\it e.g.}~charge
or mass.  Hence even if the full lower dimensional de Sitter theory would
be unstable, it is possible that a stable truncation would exist.  An
example of this is given by the truncated seven-dimensional Lagrangians
(\ref{eq:d7n2lag}) and (\ref{eq:d7n2lag2}).

While presently we have only examined multi-centered de Sitter black
holes, it would be of interest to investigate and lift other de Sitter
backgrounds to the underlying IIB$^*$ or M$^*$ theory.  In fact, such
lifted solutions may additionally be T-dualized along the time direction
so that they become conventional solutions in ten or eleven dimensions.
For the case of the lifted $D=5$ de Sitter black holes, it may be seen
that the T-dual of the rotating E4-brane solution, (\ref{eq:bhten2}),
would involve D4-branes as well as NSNS flux.

Finally, it remains an open issue whether multi-centered anti-de Sitter
black holes may be constructed in ordinary gauged supergravities without
resorting to complexification or analytic continuation.  To do so, one
would have to overcome the fact that in an ordinary supergravity, a
background preserving some fraction of the supersymmetries necessarily
admits a timelike or null Killing vector.  This is in direct
contradiction to the expectation that a multi-centered anti-de Sitter
black hole configuration would be time dependent, due to the focusing
effect of geodesics in anti-de Sitter space.  On the other hand, there
is no obstruction to the multi-centered de Sitter black holes, as the
unconventional signs in the superalgebra relax the condition of having a
timelike Killing vector.  This hints, at least, that the unconventional
de Sitter supergravities investigated here may play a crucial role in
the better understanding of de Sitter cosmologies and time dependent
backgrounds for string theory.

\section*{Acknowledgments}

This research was supported in part by the US Department of Energy under
grant DE-FG02-95ER40899.


\end{document}